# Relationship between thermodynamics and dynamics of supercooled liquids

Jeetain Mittal[1], Jeffrey R. Errington[2], & Thomas M. Truskett[1,3]

[1]*Department of Chemical Engineering, The University of Texas at Austin, Austin, TX*

[2]*Department of Chemical & Biological Engineering, University at Buffalo, The State University of New York, Buffalo, NY*

[3]*Institute for Theoretical Chemistry, The University of Texas at Austin, Austin, TX*

Diffusivity, a measure for how rapidly a fluid self-mixes, shows an intimate, but seemingly fragmented, connection to thermodynamics. On one hand, the "configurational" contribution to entropy (related to the number of mechanically-stable configurations that fluid molecules can adopt)[1] – has long been considered key for predicting supercooled liquid dynamics near the glass transition[2]. On the other hand, the excess entropy (relative to ideal gas) provides a robust scaling for the diffusivity of fluids above the freezing point[3-6]. Here we provide, to our knowledge, the first evidence that excess entropy also captures how supercooling a fluid modifies its diffusivity, suggesting that dynamics, from ideal gas to glass, is related to a single, standard thermodynamic quantity.

Several theories of the glass transition are based upon the idea that supercooled liquids vitrify when their configurational entropy vanishes[1,7-9]. Moreover, experiments[10,11] and computer simulations[12-14] reveal a quantitative link between dynamics and configurational entropy in supercooled liquids, a prediction of Adam-Gibbs theory of structural relaxation[2]. However, the configurational entropy loses relevance at high temperature, and it does not generally correlate with dynamics far above the freezing transition. As a result, it cannot provide a comprehensive description of liquid-state diffusivity. On the other hand, the excess entropy, a fundamental thermodynamic quantity which captures the correlations between particles due to their finite volumes and mutual interactions, does capture the diffusivity of equilibrium fluids[3-6]. If excess entropy turns out to also describe supercooled liquid dynamics, which is the issue we investigate here via computer simulations, then the relationship between thermodynamics and dynamics will be much simpler than previously anticipated.

We first examine the behavior of a "core-softened" fluid[15] that belongs to a larger class of model potentials known to reproduce many of liquid water's distinctive properties[15,16]. In particular, we perform simulations for a broad range of thermodynamic conditions where the model displays pronounced increases in self-diffusivity $D$ upon isothermal compression, a well-known experimental signature of supercooled liquid water's dynamics[16]. Figure 1(a, b) shows that the excess entropy $s^{ex}$ and diffusivity $D$ of this fluid have strikingly similar dependencies on density $\rho$ for a wide range of temperatures $T$. In fact, when plotted along curves of constant $\rho$ (Figure 1(c)), we find $D \propto \exp[A(\rho)s^{ex}]$, where $A(\rho)$ is a $T$-independent parameter. Fig 2 shows that this robust scaling is also exhibited by a model binary alloy[12] for conditions where it displays many of the experimental characteristics of fragile supercooled liquids[12,17,18]. This is a stringent test since this alloy has become one of the most well-characterized model glass-formers.

Adam-Gibbs theory predicts a different form of exponential relationship between $D$ and configurational entropy $s_C$, $D \propto \exp[-B(\rho)/(Ts_C)]$, where $B(\rho)$ is a $T$-independent parameter. Since Adam-Gibbs relationship can adequately describe the diffusivity for many liquids near the glass transition, it is natural to ask whether $s^{ex}$ and $-(Ts_C)^{-1}$ contain the same thermodynamic information about the supercooled fluid. Indeed, the inset of Figure 2 demonstrates that these quantities are linearly related (at constant $\rho$) for the binary alloy[12,18] over all conditions for which data is available.

The results presented here represent, to our knowledge, the first evidence that $s^{ex}$, which provides a scaling for the diffusivity of simple equilibrium fluids[3-6], also captures supercooled liquid dynamics. Moreover, since $s^{ex}$ is a standard thermodynamic quantity that can be approximated based on structural data from, e.g., scattering experiments[19], it also promises to provide the elusive link between structure and dynamics[20] of the liquid state.

We thank Srikanth Sastry, Pablo Debenedetti, and Frank Stillinger for their useful comments on an earlier version of the manuscript. Two of the authors (TMT and JRE) acknowledge the financial support of the National Science Foundation Grants No. CTS-0448721 and CTS-028772, respectively, and the Donors of the American Chemical Society Petroleum Research Fund Grants No. 41432-G5 and 43452-AC5, respectively. One of the authors (TMT) also acknowledges the support of the David and Lucile Packard and the Alfred P. Sloan Foundations. The Texas Advanced Computing Center (TACC) and the University at Buffalo Center for Computational Research provided computational resources for this study.

Email addresses: jeetain@che.utexas.edu, jerring@buffalo.edu, truskett@che.utexas.edu (corresponding author)

**Figure Captions**

FIGURE 1. (a, b) Excess entropy and diffusivity versus density obtained from molecular dynamics simulations of 1000 particles interacting via a "core-softened" potential (i.e., a Lennard-Jones potential plus a Gaussian repulsion; for details, see Reference 15). Symbols are simulation data, and curves are guides to the eye. The quantities are reported in reduced units of $T^* = k_B T/\varepsilon$, $\rho^* = \rho\sigma^3$, $D^* = D(M/\varepsilon\sigma^2)^{1/2}$, where $k_B$ is the Boltzmann constant; $T$ is the temperature; $\varepsilon$ is the energy scale of the potential; $\rho$ is the number density; $\sigma$ is the particle diameter; and $M$ is the particle mass. The excess entropy $s^{ex}$ has been calculated using transition-matrix Monte Carlo simulations[6,21]. (c) Diffusivity versus excess entropy for the data shown in (a, b) along paths of constant $\rho$ (symbols). Symbols are simulation data, and lines reflect the form $D \propto \exp[A(\rho)s^{ex}]$.

FIGURE 2. Diffusivity versus excess entropy for different density states of a binary Lennard-Jones alloy[12,18]. Excess entropy has been obtained from the semi-empirical free energy expression reported in Ref. 18, and diffusivity has been extracted from the Figure 3 in Ref. 18. The lines reflect the form $D \propto \exp[A(\rho)s^{ex}]$. The inset shows the linear dependence between excess entropy and the inverse product of temperature and configurational entropy, $1/Ts_C$, the latter extracted from Figure 2 in Ref. 12. The quantities are reported in the same reduced units as in Figure 1.

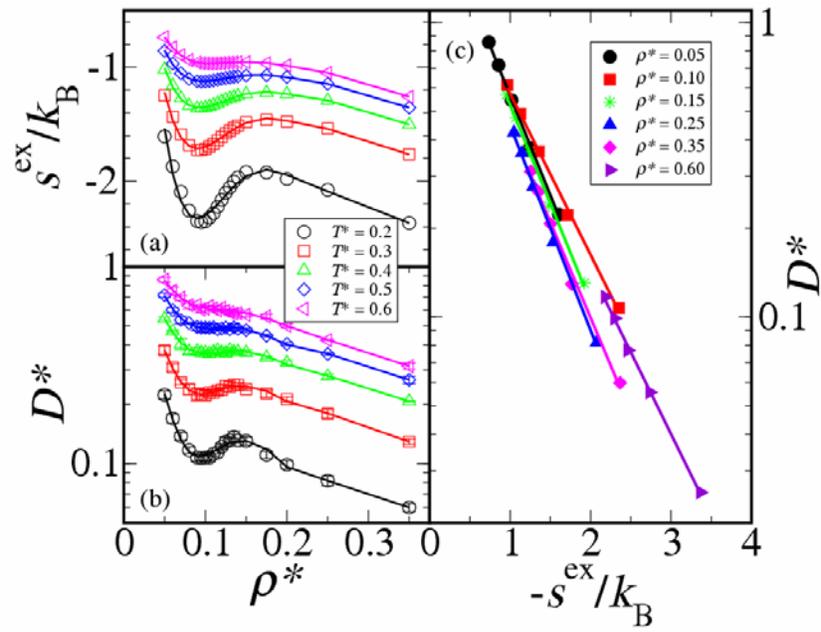

Figure 1. Mittal, Errington, & Truskett (2006)

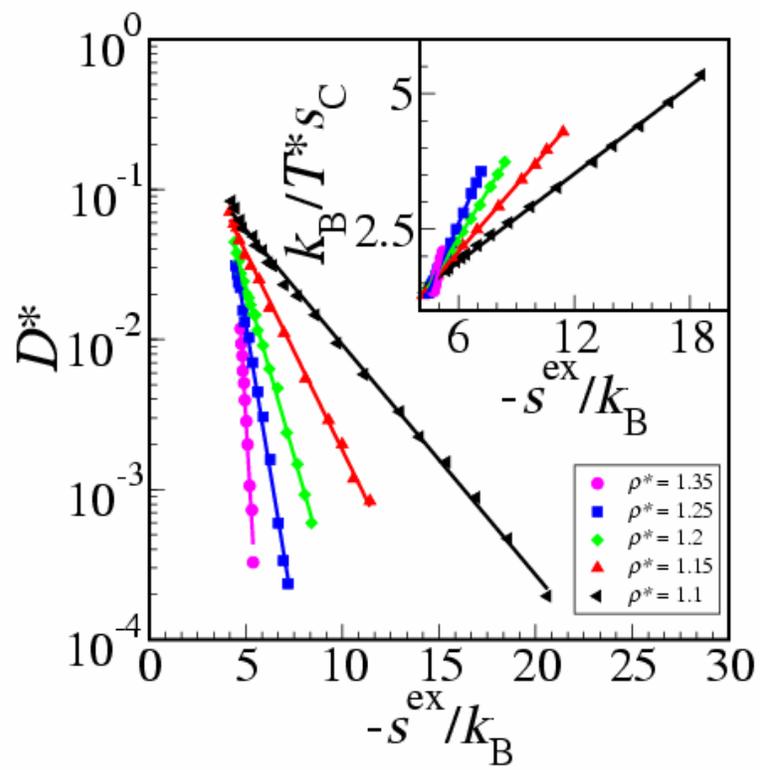

Figure 2. Mittal, Errington, & Truskett (2006)